# Finite Mixtures of Canonical Fundamental Skew $t$-Distributions

Sharon X. Lee, Geoffrey J. McLachlan

Department of Mathematics, the University of Queensland,

Brisbane, Australia

**Abstract**

This is an extended version of the paper Lee and McLachlan (2014b) with simulations and applications added. This paper introduces a finite mixture of canonical fundamental skew $t$ (CFUST) distributions for a model-based approach to clustering where the clusters are asymmetric and possibly long-tailed (Lee and McLachlan, 2014b). The family of CFUST distributions includes the restricted multivariate skew $t$ (rMST) and unrestricted multivariate skew $t$ (uMST) distributions as special cases. In recent years, a few versions of the multivariate skew $t$ (MST) model have been put forward, together with various EM-type algorithms for parameter estimation. These formulations adopted either a restricted or unrestricted characterization for their MST densities. In this paper, we examine a natural generalization of these developments, employing the CFUST distribution as the parametric family for the component distributions, and point out that the restricted and unrestricted characterizations can be unified under this general formulation. We show that an exact implementation of the EM algorithm can be achieved for the CFUST distribution and mixtures of this distribution, and present some new analytical results for a conditional expectation involved in the E-step.

## 1 Introduction

To date, various parametric families of densities have been proposed in the literature to represent the multivariate skew $t$ (MST) model, all having either a restricted or unrestricted characterization according to the classification scheme presented in Lee and McLachlan (2013a). As stressed by Lee and McLachlan (2013a), their use of "restricted" refers to restrictions on the random vector in the (convolution-type) stochastic definition of the skew distribution. It is not a restriction on the parameter space, and so a restricted form of a skew distribution is not necessarily nested within its corresponding "unrestricted" form. In particular, the restricted MST (rMST) mixture model (or its equivalent variants by reparameterization) have been examined in Pyne et al. (2009), Cabral et al. (2012) and Vrbik and McNicholas (2012), while Lin (2010), and Lee and McLachlan (2011) focused on the unrestricted MST (uMST) mixture model adopting the characterization proposed by Sahu et al. (2003). In addition, Murray et al. (2013) recently considered an alternative construction that also adopted the term 'skew $t$-distribution', although referring to a rather different form more commonly known as the generalized hyperbolic skew $t$ (GHST) distribution. The GHST distribution is a restricted type of skew distribution in the sense that the skewing function is univariate. It differs from the rMST distribution in a number of ways, such as different behaviour in its tails with one polynomial and the other exponential (Aas and Haff, 2005). Also, it does not become a skew normal distribution as a limiting case (Lee and Poon, 2011).

In this paper, we consider a broader class of distributions that contains both the rMST and uMST distributions. The skewing function for the densities in this family is specified by a $q$-dimensional $t$-distribution function, and the skewness parameter $\boldsymbol{\Delta}$ is a general $p$ by $q$ matrix. This formulation coincides with a special case of the unified skew $t$ (SUT) distribution (Arellano-Valle and Azzalini, 2006) and the fundamental skew $t$ (FUST) distribution (Arellano-Valle and Genton, 2005), and is known as a location-scale variant of the canonical fundamental skew $t$ (CFUST) distribution. An appealing feature of this characterization is that it formally incorporates both the restricted and unrestricted forms of the MST distribution.

As discussed in Lee and McLachlan (2014a), while exact implementation of the Expectation-Maximization (EM) algorithm has been achievable for restricted characterizations of the component skew $t$-distributions, parameter estimation for the unrestricted and more general characterizations is substantially more intricate. In previous work on mixtures of unrestricted skew $t$-distributions using the characterization of Sahu et al. (2003) (a submodel of the CFUST distribution considered in this paper), Lin (2010) resorted to Monte Carlo methods for the computation of intractable conditional expectations involved in the E-step, while Lee and McLachlan (2014a) adopted the OSL (one-step late) approach to evaluating one of these difficult conditional expectations. In contrast, the E-step is straightforward for the rMST distribution and so there exist a number of implementations of the EM algorithm (see Lee and McLachlan (2014a) for further discussion).

This paper presents an EM algorithm for fitting the FM-CFUST model. Under this model, the aforementioned restricted and unrestricted skew $t$-mixture models can be obtained as special cases. As such, the expressions on the E-step for the FM-uMST model can be extended in a straightforward manner to the FM-CFUST model. The M-step for the latter model requires only slight modifications to that for the FM-uMST model. In contrast to the approach of Lin (2010) which relies on computationally intensive numerical approximations, our development was motivated by the desire for an exact implementation the EM algorithm using the truncated moments approach (Lee and McLachlan, 2014a, Ho et al., 2012), leading to much faster and more accurate results. Also, we derive a new exact expression for the remaining E-step conditional expectation that had previously relied on approximation or numerical methods.

The methodology is illustrated on the analysis of five simulated and two real datasets. In the five simulated sets, we explore the capability of the CFUST model in modelling data generated from both the restricted and unrestricted distributions, as well as some multivariate skew data for which the latter two models have some difficulty in handling. In Section 5, the performance of the FM-CFUST model is further compared to these models in clustering two real datasets. With the availability of software for the fitting of the FM-CFUST model, users now have the option of letting the data implicitly decide as to which model is appropriate through the estimation of the component-matrices of skewness parameters in the convolution-type formulation of the skew $t$-mixture model. Alternatively, the user can explicitly decide between the three models using a model selection criterion, such as BIC (Schwarz, 1978).

## 2  The CFUST distribution

The (multivariate) canonical fundamental skew $t$-distribution (CFUST) was introduced as a canonical version (special case) of the fundamental skew $t$ (FUST) by Arellano-Valle and Genton (2005). A location-scale variant of the CFUST distribution can be characterized as follows.

Let $U_1$ be $p$-dimensional random vector and $U_0$ a $q$-dimensional random vector. Suppose that, conditional on a value $w$ of a gamma random variable $W$ distributed as gamma$(\frac{\nu}{2}, \frac{\nu}{2})$, the joint distribution of $U_0$ and $U_1$ is given by

$$\begin{bmatrix} U_0 \\ U_1 \end{bmatrix} \sim N_{q+p}\left(\begin{bmatrix} 0 \\ 0 \end{bmatrix}, \frac{1}{w}\begin{bmatrix} I_q & 0 \\ 0 & \Sigma \end{bmatrix}\right), \quad (1)$$

where $\nu$ is a scalar, $I_q$ denotes the $q \times q$ identity matrix, $\Sigma$ is a positive definite scale matrix, $\Delta$ is a $p \times q$ matrix, and $0$ is a vector/ matrix of zeros with appropriate dimensions. Then

$$Y = \mu + \Delta |U_0| + U_1 \quad (2)$$

follows the CFUST distribution, whose density is defined by

$$\begin{aligned} f(y; \mu, \Sigma, \Delta, \nu) &= 2^q\, t_p(y; \mu, \Omega, \nu) \\ &\quad T_q\left(c(y)\sqrt{\frac{\nu+p}{\nu+d(y)}}; 0, \Lambda, \nu+p\right), \end{aligned} \quad (3)$$

where

$$\begin{aligned} \Omega &= \Sigma + \Delta\Delta^T, \\ c(y) &= \Delta^T \Omega^{-1}(y - \mu), \\ \Lambda &= I_p - \Delta^T \Omega^{-1}\Delta, \\ d(y) &= (y - \mu)^T \Omega^{-1}(y - \mu). \end{aligned}$$

Here, we let $t_p(y; \mu, \Omega, \nu)$ denote the $p$-dimensional $t$-distribution with location parameter $\mu$, scale matrix $\Omega$, and degrees of freedom $\nu$, and $T_q(.)$ is the $q$-dimensional (cumulative) $t$-distribution function. In the above, $|U_0|$ denotes the vector whose $i$th element is the magnitude of the $i$th element of the vector $U_0$.

## 2.1 The unrestricted characterization as a special case

As mentioned previously, the unrestricted MST (uMST) distribution (the characterization of Sahu et al. (2003)) corresponds to the CFUST distribution with $q = p$ and $\Delta = \text{diag}(\delta)$. Hence, we can write $\Lambda = I_p - \Delta\Omega^{-1}\Delta$, $c(y) = \Delta\Omega^{-1}(y - \mu)$, and $\Omega = \Sigma + \Delta^2$, and the uMST density is given by

$$\begin{aligned} f(y; \mu, \Sigma, \delta, \nu) &= 2^p\, t_p(y; \mu, \Omega, \nu) \\ &\quad T_p\left(c(y)\sqrt{\frac{\nu+p}{\nu+d(y)}}; 0, \Lambda, \nu+p\right). \end{aligned} \quad (4)$$

The uMST distribution has the same stochastic representation as given by (1) and (2) with $q=p$, and $\Delta$ taken to be diagonal.

## 2.2 The restricted characterization as a special case

In the restricted characterization of the MST distribution, the convolution-type stochastic representation is given by

$$Y = \mu + \delta |U_0| + U_1, \quad (5)$$

where

$$\begin{bmatrix} U_0 \\ \boldsymbol{U}_1 \end{bmatrix} \sim N_{1+p}\left(\begin{bmatrix} 0 \\ \boldsymbol{0} \end{bmatrix}, \frac{1}{w}\begin{bmatrix} 1 & \boldsymbol{0} \\ \boldsymbol{0} & \boldsymbol{\Sigma} \end{bmatrix}\right). \tag{6}$$

That is, the restricted formulation is given by imposing the restriction that the elements of the vector $\boldsymbol{U}_0$ are all equal to $U_0$ in (2); hence the use of "restricted" in the classification scheme of Lee and McLachlan (2013a). It follows that the density of the rMST distribution is given by

$$\begin{aligned} f(\boldsymbol{y}) &= 2t_p(\boldsymbol{y}; \boldsymbol{\mu}, \boldsymbol{\Omega}, \nu) \\ & \quad T_1\left(\boldsymbol{\delta}^T\boldsymbol{\Omega}^{-1}(\boldsymbol{y}-\boldsymbol{\mu}), \left(\frac{\nu+d(\boldsymbol{y})}{\nu+p}\right)\lambda, \nu+p\right), \end{aligned} \tag{7}$$

where $\lambda = 1 - \boldsymbol{\delta}^T\boldsymbol{\Omega}^{-1}\boldsymbol{\delta}$.

It can be seen from (7) that the restricted density has a univariate skewing function as opposed to the multivariate version in (4) for the unrestricted skew t-density. On directly comparing (5) and (2), it can be observed that the skewing effect of $\boldsymbol{\delta}|U_0|$ in the rMST distribution can be achieved by the CFUST formulation in a number of ways, including

(i) taking $\boldsymbol{\Delta} = \text{diag}(\boldsymbol{\delta})$ as in the unrestricted case, and setting $\boldsymbol{U}_0 = U_0 \boldsymbol{1}_p$ in (2), where $\boldsymbol{1}_p$ denotes a $p$-dimensional vector with elements one;

(ii) constraining the skewness matrix $\boldsymbol{\Delta}$ to be a $p \times p$ zero matrix except for one column which is given by $\boldsymbol{\delta}$, and setting the corresponding element in $\boldsymbol{U}_0$ to be $U_0$; or

(iii) setting $q = 1$.

As a corollary, it can be seen that the special case of $q = p$ of the CFUST distribution encompasses both the rMST and uMST distributions through structural constraints on the skewness parameter. This is given by taking

$$\boldsymbol{\Delta} = \begin{bmatrix} \delta_1 & 0 & \ldots & 0 \\ \delta_2 & 0 & \ldots & 0 \\ \vdots & \vdots & \ddots & \vdots \\ \delta_p & 0 & \ldots & 0 \end{bmatrix} \text{ and } \boldsymbol{\Delta} = \begin{bmatrix} \delta_1 & 0 & \ldots & 0 \\ 0 & \delta_2 & \ldots & 0 \\ \vdots & \vdots & \ddots & \vdots \\ 0 & 0 & \ldots & \delta_p \end{bmatrix} \tag{8}$$

for the rMST and uMST distributions respectively, where $\boldsymbol{\delta} = (\delta_1, \delta_2, \ldots, \delta_p)^T$. Note that for the rMST case, $\boldsymbol{\delta}$ can be placed in any one of the columns of $\boldsymbol{\Delta}$, not necessarily the first column, as given in (8).

## 3 Parameter estimation via the EM algorithm

We are interested in finite mixtures with component densities given by (3). A $g$-component mixture of CFUST distributions has density

$$f(\boldsymbol{y}; \boldsymbol{\Psi}) = \sum_{h=1}^{g} \pi_h f(\boldsymbol{y}; \boldsymbol{\mu}_h, \boldsymbol{\Sigma}_h, \boldsymbol{\Delta}_h, \nu_h), \tag{9}$$

where $\pi_h$ ($h = 1, \ldots, g$) are the mixing proportions and $f(.)$ denotes a CFUST density given by (3). We shall adopt the acronym FM-CFUST for (9). The vector $\boldsymbol{\Psi} = (\pi_1, \ldots, \pi_{g-1}, \boldsymbol{\theta}_1^T, \ldots, \boldsymbol{\theta}_g^T)$ contains all the unknown parameters of the model, with $\boldsymbol{\theta}_h$ containing the elements of $\boldsymbol{\mu}_h$ and $\boldsymbol{\delta}_h$, the distinct elements of $\boldsymbol{\Sigma}_h$, and $\nu_h$.

Similar to the rMST and uMST distributions, the CFUST admits a convenient hierarchical representation that greatly facilitates parameter estimation via the EM algorithm:

$$\begin{aligned}
\boldsymbol{Y} \mid \boldsymbol{u}, w &\sim N_p\left(\boldsymbol{\mu} + \boldsymbol{\Delta}\boldsymbol{u}, \frac{1}{w}\boldsymbol{\Sigma}\right), \\
\boldsymbol{U} \mid w &\sim HN_q\left(\boldsymbol{0}, \frac{1}{w}\boldsymbol{I}_q\right), \\
W &\sim \text{gamma}\left(\frac{\nu}{2}, \frac{\nu}{2}\right),
\end{aligned} \quad (10)$$

where $N_p(\boldsymbol{\mu}, \boldsymbol{\Sigma})$ denotes the multivariate normal distribution and $HN_q(0, \boldsymbol{\Sigma})$ represents the $q$-dimensional half-normal distribution with mean $\boldsymbol{0}$ and scale matrix $\boldsymbol{\Sigma}$. The EM algorithm for the FM-CFUST model then proceeds in a similar way as for the FM-uMST model as described in Lee and McLachlan (2014a).

## 3.1 E-Step

On the $(k+1)$th iteration of the E-step in the application of the EM algorithm to the fitting of the FM-CFUST model, the following conditional expectations need to be calculated:

$$\begin{aligned}
z_{hj}^{(k)} &= E_{\boldsymbol{\Psi}^{(k)}}\left[z_{hj} = 1 \mid \boldsymbol{y}_j\right], & (11) \\
w_{hj}^{(k)} &= E_{\boldsymbol{\Psi}^{(k)}}\left[w_{hj} \mid \boldsymbol{y}_j, z_{hj} = 1\right], & (12) \\
e_{1hj}^{(k)} &= E_{\boldsymbol{\Psi}^{(k)}}\left[\log(w_{hj}) \mid \boldsymbol{y}_j, z_{hj} = 1\right], & (13) \\
\boldsymbol{e}_{2hj}^{(k)} &= E_{\boldsymbol{\Psi}^{(k)}}\left[w_{hj}\boldsymbol{u}_{hj} \mid \boldsymbol{y}_j, z_{hj} = 1\right], & (14) \\
\boldsymbol{e}_{3hj}^{(k)} &= E_{\boldsymbol{\Psi}^{(k)}}\left[w_{hj}\boldsymbol{u}_{hj}\boldsymbol{u}_{hj}^T \mid, \boldsymbol{y}_j, z_{hj} = 1\right]. & (15)
\end{aligned}$$

These expectations can be expressed as

$$\begin{aligned}
z_{hj}^{(k)} &= \frac{\pi_h f(\boldsymbol{y}_j; \boldsymbol{\mu}_h^{(k)}, \boldsymbol{\Sigma}_h^{(k)}, \boldsymbol{\delta}_h^{(k)}, \nu_h^{(k)})}{f(\boldsymbol{y}_j; \boldsymbol{\Psi}^{(k)})}, \\
w_{hj}^{(k)} &= \left(\frac{\nu_h^{(k)} + p}{\nu_h^{(k)} + d_h^{(k)}(\boldsymbol{y}_j)}\right) \\
&\quad \frac{T_q\left(\boldsymbol{c}_h^{(k)}\sqrt{\frac{\nu_h^{(k)}+p+2}{\nu_h^{(k)}d_h^{(k)}(\boldsymbol{y}_j)}}; \boldsymbol{0}, \boldsymbol{\Lambda}_h^{(k)}, \nu_h^{(k)} + p + 2\right)}{T_q\left(\boldsymbol{c}_h^{(k)}\sqrt{\frac{\nu_h^{(k)}+p}{\nu_h^{(k)}d_h^{(k)}(\boldsymbol{y}_j)}}; \boldsymbol{0}, \boldsymbol{\Lambda}_h^{(k)}, \nu_h^{(k)} + p\right)},
\end{aligned} \quad (16)$$

$$e^{(k)}_{1hj} = w^{(k)}_{hj} - \log\left(\frac{\nu^{(k)}_h + d^{(k)}_h(\boldsymbol{y}_j)}{2}\right)$$
$$+ \left(\frac{\nu^{(k)}_h + p}{\nu^{(k)}_h + d^{(k)}_h(\boldsymbol{y}_j)}\right) - \psi\left(\frac{\nu^{(k)}_h + p}{2}\right), \tag{17}$$

$$\boldsymbol{e}^{(k)}_{2,hj} = w^{(k)}_{hj} E_{\boldsymbol{\Psi}^{(k)}}[\boldsymbol{u}_{hj} \mid \boldsymbol{y}_j],$$
$$\boldsymbol{e}^{(k)}_{3hj} = w^{(k)}_{hj} E_{\boldsymbol{\Psi}^{(k)}}[\boldsymbol{u}_{hj}\boldsymbol{u}^T_{hj} \mid \boldsymbol{y}_j], \tag{18}$$

where $\boldsymbol{U}_{hj} \mid \boldsymbol{y}_j$ has a $q$-dimensional truncated $t$-distribution given by

$$\boldsymbol{U}_{hj} \mid \boldsymbol{y}_j \sim tt_q\left(\boldsymbol{c}^{(k)}_h, \left(\frac{\nu^{(k)}_h + d_h(\boldsymbol{y}_j)}{\nu^{(k)}_h + p + 2}\right)\boldsymbol{\Lambda}^{(k)}_h,\right.$$
$$\left.\nu^{(k)}_h + p + 2; \mathbb{R}^+\right).$$

Note that (17) is obtained using an OSL-type EM algorithm. The derivations for these results are analogous to Lee and McLachlan (2014a). Alternatively, a series-based truncation approach, which exploits an exact representation of this conditional expectation, can be used in place of (17). It is to be presented in Section 3.4.

## 3.2 M-Step

On the $(k+1)$th iteration of the M-step, the model parameters are updated to give

$$\pi^{(k+1)}_h = \frac{1}{n}\sum_{j=1}^n z^{(k)}_{hj},$$

$$\boldsymbol{\mu}^{(k+1)}_h = \frac{\sum_{j=1}^n z^{(k)}_{hj} w^{(k)}_{hj} \boldsymbol{y}_j - \boldsymbol{\Delta}^{(k)}_h \sum_{j=1}^n z^{(k)}_{hj} \boldsymbol{e}^{(k)}_{2hj}}{\sum_{j=1^n} z^{(k)}_{hj} w^{(k)}_{hj}},$$

$$\boldsymbol{\Delta}^{(k+1)}_h = \left[\sum_{j=1}^n z^{(k)}_{hj}\left(\boldsymbol{y}_j - \boldsymbol{\mu}^{(k+1)}_h\right)\boldsymbol{e}^{(k)T}_{2hj}\right]$$
$$\left[\sum_{j=1}^n z^{(k)}_{hj} \boldsymbol{e}^{(k)}_{3hj}\right]^{-1}, \tag{19}$$

$$\boldsymbol{\Sigma}^{(k+1)}_h = \left\{\sum_{j=1}^n z^{(k)}_{hj}\left[w^{(k)}_{hj}\left(\boldsymbol{y}_j - \boldsymbol{\mu}^{(k+1)}_h\right)\left(\boldsymbol{y}_j - \boldsymbol{\mu}^{(k+1)}_h\right)^T\right.\right.$$
$$-\boldsymbol{\Delta}^{(k+1)}_h \boldsymbol{e}^{(k)}_{2hj}\left(\boldsymbol{y}_j - \boldsymbol{\mu}^{(k+1)}_h\right)^T$$
$$-\left(\boldsymbol{y}_j - \boldsymbol{\mu}^{(k+1)}_h\right)\boldsymbol{e}^{(k)T}_{2hj}\boldsymbol{\Delta}^{(k+1)T}_h$$
$$\left.\left.+\boldsymbol{\Delta}^{(k+1)}_h \boldsymbol{e}^{(k)T}_{3hj}\boldsymbol{\Delta}^{(k+1)T}_h\right]\right\}\left[\sum_{j=1}^n z^{(k)}_{hj}\right]^{-1}. \tag{20}$$

An update of the degrees of freedom $\nu_h$ is obtained by solving the following equation for

$\nu_h^{(k+1)}$,

$$0 = \left(\sum_{h=1}^{n} z_{hj}^{(k)}\right) \left[\log\left(\frac{\nu_h^{(k+1)}}{2}\right) - \psi\left(\frac{\nu_h^{(k+1)}}{2}\right) + 1\right]$$
$$- \left(\sum_{j=1}^{n} e_{1hj}^{(k)} - w_{hj}^{(k)}\right),$$

where $\psi(\cdot)$ denotes the digamma function.

### 3.3 The rMST and uMST as special cases

As pointed out previously, the rMST and uMST distributions can be obtained as special cases of the CFUST distribution by imposing structural constraints on the matrix $\boldsymbol{\Delta}$ of skewness parameters. In the case of the uMST distribution, the component skewness matrices $\boldsymbol{\Delta}_h$ are taken to be diagonal; that is, $\boldsymbol{\Delta}_h = \text{diag}(\boldsymbol{\delta}_h)$. These elements are estimated as

$$\boldsymbol{\delta}_h^{(k+1)} = \left(\boldsymbol{\Sigma}_h^{(k)^{-1}} \circ \sum_{j=1}^{n} \tau_{hj}^{(k)} \boldsymbol{e}_{4,hj}^{(k)}\right)^{-1}$$
$$\text{diag}\left[\boldsymbol{\Sigma}_h^{(k)^{-1}} \sum_{j=1}^{n} \tau_{hj}^{(k)} \left(\boldsymbol{y}_j - \boldsymbol{\mu}_h^{(k+1)}\right) \boldsymbol{e}_{3,hj}^{(k)^T}\right], \quad (21)$$

where $\circ$ denotes the elementwise product.

Under the constraint that $\boldsymbol{\Delta}_h$ is zero except for the first column $\boldsymbol{\delta}_h$, that is, $\boldsymbol{\Delta} = [\boldsymbol{\delta}\, \boldsymbol{0} \ldots \boldsymbol{0}]$, (19) reduces to estimating

$$\boldsymbol{\delta}_h^{(k+1)} = \frac{\sum_{j=1}^{n} \tau_{hj}^{(k)} [\boldsymbol{e}_{3,hj}^{(k)}]_1 (\boldsymbol{y}_j - \boldsymbol{\mu}_h^{(k+1)})}{\sum_{j=1}^{n} \tau_{hj}^{(k)} [\boldsymbol{e}_{4,hj}^{(k)}]_{11}}, \quad (22)$$

where $[\boldsymbol{e}_{3,hj}^{(k)}]_1$ and $[\boldsymbol{e}_{4,hj}^{(k)}]_{11}$ denote the first element of $\boldsymbol{e}_{3,hj}^{(k)}$ and $\boldsymbol{e}_{4,jh}^{(k)}$, respectively. On comparing the above with the corresponding expression for the rMST mixture model (see, for example, equation (9) from Wang et al. (2009)), it can observed that the two expressions are indeed equivalent.

### 3.4 New results on $e_{1,hj}^{(k)}$: the truncated series approach

We now present the derivation of a new exact expression for $e_{1,hj}^{(k)}$, expressed in terms of an infinite series involving multivariate $t$-distribution functions. By the law of iterated expectations,

$$\begin{aligned} e_{1,hj}^{(k)} &= E_{\boldsymbol{\Psi}^{(k)}}\left[\log(W_j) \mid \boldsymbol{y}_j\right] \\ &= E_{\boldsymbol{\Psi}^{(k)}}\left[E_{\boldsymbol{\Psi}^{(k)}}\left[\log(W_j) \mid \boldsymbol{u}_j, \boldsymbol{y}_j\right] \mid \boldsymbol{y}_j\right] \\ &= \psi\left(\frac{\nu_h^{(k)} + p}{2}\right) - \log\left(\frac{\nu_h^{(k)} + d_h^{(k)}(\boldsymbol{y}_j)}{2}\right) \\ &\quad - E_{\boldsymbol{\Psi}^{(k)}}\left[\log\left(1 + \frac{(\boldsymbol{u}_j - \boldsymbol{c}_{hj}^{(k)})^T \boldsymbol{\Lambda}_h^{(k)^{-1}} (\boldsymbol{u}_j - \boldsymbol{c}_{hj}^{(k)})}{\nu_h^{(k)} + d_h^{(k)}(\boldsymbol{y}_j)}\right)\right]. \end{aligned}$$
(23)

To calculate the last term in (23), note that the conditional density of $\boldsymbol{u}_j$ given $\boldsymbol{y}_j$ can be written as

$$f(\boldsymbol{u}_j \mid \boldsymbol{y}_j) = \frac{t_q\left(\boldsymbol{u}_j; \boldsymbol{c}_{hj}^{(k)}, \left(\frac{\nu_h^{(k)}+d_h^{(k)}(\boldsymbol{y}_j)}{\nu_h^{(k)}+p}\right)\boldsymbol{\Lambda}_h^{(k)}, \nu_h^{(k)}+p\right)}{T_q\left(\boldsymbol{c}_{hj}^{(k)}; \boldsymbol{0}, \left(\frac{\nu_h^{(k)}+d_h^{(k)}(\boldsymbol{y}_j)}{\nu_h^{(k)}+p}\right)\boldsymbol{\Lambda}_h^{(k)}, \nu_h^{(k)}+p\right)}.$$

After some algebraic manipulations, it follows that the expectation in (23) can be reduced to a simple closed-form expression (see equation (76) from Lee and McLachlan (2014a)), except for a term involving an integral given by

$$\begin{aligned}
S_{1,hj}^{(k)} &= \int_{\boldsymbol{0}}^{\infty} \log\left(1 + \frac{(\boldsymbol{u}_j - \boldsymbol{c}_{hj}^{(k)})^T \boldsymbol{\Lambda}_h^{(k)-1}(\boldsymbol{u}_j - \boldsymbol{c}_{hj}^{(k)})}{\nu_h^{(k)}+d_h^{(k)}(\boldsymbol{y}_j)}\right) \\
&\quad t_q\left(\boldsymbol{u}_j; \boldsymbol{c}_{hj}^{(k)}, \left(\frac{\nu_h^{(k)}+d_h^{(k)}(\boldsymbol{y}_j)}{\nu_h^{(k)}+p}\right)\boldsymbol{\Lambda}_h^{(k)}, \nu_h^{(k)}+p\right) d\boldsymbol{u}_j \\
&= \sum_{r=1}^{\infty}\sum_{s=0}^{r} \frac{(-1)^{2r-s-1}}{r} \binom{r}{s} \\
&\quad \int_{\boldsymbol{0}}^{\infty}\left(1 + \frac{(\boldsymbol{u}_j - \boldsymbol{c}_{hj}^{(k)})^T \boldsymbol{\Lambda}_h^{(k)-1}(\boldsymbol{u}_j - \boldsymbol{c}_{hj}^{(k)})}{\nu_h^{(k)}+d_h^{(k)}(\boldsymbol{y}_j)}\right)^{-s} \\
&\quad t_q\left(\boldsymbol{u}_j; \boldsymbol{c}_{hj}^{(k)}, \left(\frac{\nu_h^{(k)}+d_h^{(k)}(\boldsymbol{y}_j)}{\nu_h^{(k)}+p}\right)\boldsymbol{\Lambda}_h^{(k)}, \nu_h^{(k)}+p\right) d\boldsymbol{u}_j \\
&\hspace{10cm}(24) \\
&= \sum_{r=1}^{\infty}\sum_{s=0}^{r} \frac{(-1)^{2r-s-1}}{r} \binom{r}{s} \\
&\quad \frac{\Gamma\left(\frac{\nu_h^{(s)}+p}{2}+s\right)}{\Gamma\left(\frac{\nu_h^{(k)}+p}{2}\right)} \frac{\Gamma\left(\frac{\nu_h^{(k)}+p+q}{2}\right)}{\Gamma\left(\frac{\nu_h^{(k)}+p+q}{2}+s\right)} \\
&\quad \int_{\boldsymbol{0}}^{\infty} t_q\left(\boldsymbol{u}_j; \boldsymbol{c}_{hj}^{(k)}, \left(\frac{\nu_h^{(k)}+d_h^{(k)}(\boldsymbol{y}_j)}{\nu_h^{(k)}+p+2s}\right)\boldsymbol{\Lambda}_h^{(k)},\right. \\
&\hspace{5cm}\left. \nu_h^{(k)}+p+2s\right) d\boldsymbol{u}_j \hspace{2cm}(25)
\end{aligned}$$

$$
\begin{aligned}
= \;& \frac{\Gamma\left(\frac{\nu_h^{(k)}+p+q}{2}\right)}{\Gamma\left(\frac{\nu_h^{(k)}+p}{2}\right)} \sum_{r=1}^{\infty}\sum_{s=0}^{r} \frac{(-1)^{2r-s-1}}{r}\binom{r}{s} \\
& \frac{\Gamma\left(\frac{\nu_h^{(s)}+p}{2}+s\right)}{\Gamma\left(\frac{\nu_h^{(k)}+p+q}{2}+s\right)} \\
& T_q\left(\boldsymbol{c}_{hj}^{(k)};\boldsymbol{0},\left(\frac{\nu_h^{(k)}+d_h^{(k)}(\boldsymbol{y}_j)}{\nu_h^{(k)}+p+2s}\right)\boldsymbol{\Lambda}_h^{(k)},\right. \\
& \left. \nu_h^{(k)}+p+2s\right).
\end{aligned} \tag{26}
$$

In (24) above, we have applied a Taylor expansion of $\log(x)$ for $0 < x < 2$ and the binomial expansion formula to obtain a series expression for $\log(x)$, given by

$$\log(x) = \sum_{r=1}^{\infty}\sum_{s=0}^{r} \frac{(-1)^{2r-s-1}}{r}\binom{r}{s} x^s, \quad \text{for } 0 < x < 2.$$

Note that the term inside the logarithm in (25) is always greater than one. Hence, its inverse lies between 0 and 1, and we can apply the above series representation. In practice, $S_{1,hj}^{(k)}$ can be approximated by a (small) finite number of terms.

It follows that $e_{1,hj}^{(k)}$ is given by

$$
\begin{aligned}
e_{1,hj}^{(k)} =\;& \psi\left(\frac{\nu_h^{(k)}+p}{2}\right) - \log\left(\frac{\nu_h^{(k)}+d_h^{(k)}(\boldsymbol{y}_j)}{2}\right) \\
& + \frac{\Gamma\left(\frac{\nu_h^{(k)}+p+q}{2}\right)}{\Gamma\left(\frac{\nu_h^{(k)}+p}{2}\right)} \\
& T_q^{-1}\left(\boldsymbol{c}_{hj}^{(k)};\boldsymbol{0},\left(\frac{\nu_h^{(k)}+d_h^{(k)}(\boldsymbol{y}_j)}{\nu_h^{(k)}+p}\right)\boldsymbol{\Lambda}_h^{(k)},\nu_h^{(k)}+p\right) \\
& \sum_{r=1}^{\infty}\sum_{s=0}^{r}\frac{(-1)^{2r-s-1}}{r}\binom{r}{s}\frac{\Gamma\left(\frac{\nu_h^{(s)}+p}{2}+s\right)}{\Gamma\left(\frac{\nu_h^{(k)}+p+q}{2}+s\right)} \\
& T_q\left(\boldsymbol{c}_{hj}^{(k)};\boldsymbol{0},\left(\frac{\nu_h^{(k)}+d_h^{(k)}(\boldsymbol{y}_j)}{\nu_h^{(k)}+p+2s}\right)\boldsymbol{\Lambda}_h^{(k)},\nu_h^{(k)}+p+2s\right).
\end{aligned} \tag{27}
$$

## 3.5 Starting Values

On the initialization of parameters for the FM-CFUST model, one may consider a number of different starting strategies based on the fitted results of its nested models, such as the restricted skew $t$-mixture model, the unrestricted skew $t$-mixture model, and (ordinary) $t$-mixture models. In addition, one can also initialize the parameters of the FM-CFUST model based on an initial partition of the data given by the above models, as well as that given by random and $k$-means based starts.

In our approach to working with these skew $t$-mixture models, we usually fit mixtures of restricted skew $t$-distributions in the first instance given that they can be fitted much more quickly than the unrestricted version. In this case, the initial cluster labels are given by a $k$-means clustering, and the initial values of $\boldsymbol{\mu}_h$ and $\boldsymbol{\Sigma}_h$ are specified to be the sample mean and sample covariance matrix, respectively, of the corresponding cluster. The skewness parameters $\boldsymbol{\delta}_h$ are initialized based on the sample skewness of the $h$th cluster. The initial degrees of freedom $\nu_h^{(0)}$ takes a small fixed value. In practice, the EM algorithm is initialized with multiple different initial partitions, and the set of parameters corresponding to the highest log likelihood value is used.

In proceeding then to fit the FM-uMST models, the clustering provided by the restricted model can be used as one of the initial partitions of the data. This is in addition to using random, $k$-means, and (ordinary) $t$-mixtures based starts; see Lee and McLachlan (2014a) for a description of an initialization of parameters based on a given initial clustering.

Then to fit the more general FM-CFUST distribution in the case where $\boldsymbol{\Delta}$ is taken to be a $p \times p$ matrix, the partitions provided by the FM-rMST and FM-uMST models can be used. In the former case, $\boldsymbol{\Delta}$ is taken to be diagonal on the first iteration of the M-step, using the estimate of $\boldsymbol{\Delta}_h$ for the $h$th component provided by the FM-uMST model. In the event that we are unable to find a local maximum that is greater than the value of the local maximum providing the ML solution under the FM-uMST model, we take the ML solution to be that provided by the FM-uMST model. Similarly, we take the ML solution to be that provided by the FM-rMST model in those situations where the local maximum providing the ML solution for the FM-rMST model is greater than any local maximum found under the FM-uMST and FM-CFUST models.

# 4 Analysis of Simulated Data

## 4.1 Simulation of a mixture of rMST and uMST observations

To highlight the differences between these models, we consider an artificial dataset with observations generated from a two-component FM-CFUST distribution, where one component has a restricted skew $t$-distribution and the other has an unrestricted skew $t$-distribution. The simulated dataset contains 1000 observations, with 500 observations from each component. The FM-CFUST, FM-uMST, and FM-rMST models are fitted to these simulated data with $g$ predefined as 2.

Figure 1 shows the contours of the fitted models to the simulated data. Of interest here is to compare the fits given by these models, in particular, the shape of each component of the fitted distributions, as well as their clustering performance. Focusing first on the lower cluster of data in Figure 1, corresponding to those data points generated from the rMST component (shown as red dots in Figure 1), it can be observed that the fits given by the FM-rMST and FM-CFUST models are quite similar. It can also be observed that they adapt to the shape of this cluster better than the fit given by the FM-uMST model. This is not surprising considering that the data were generated from a rMST distribution. It is encouraging that the FM-CFUST model without prior knowledge of any constraints on its component matrices $\boldsymbol{\Delta}_h$ of skewness parameters is able to model the data accordingly. For the FM-uMST model, it can be argued from Figure 1(c) that the fit given by the FM-uMST model is not unreasonable. However, it can impact adversely on its clustering performance depending on the orientation of the clusters.

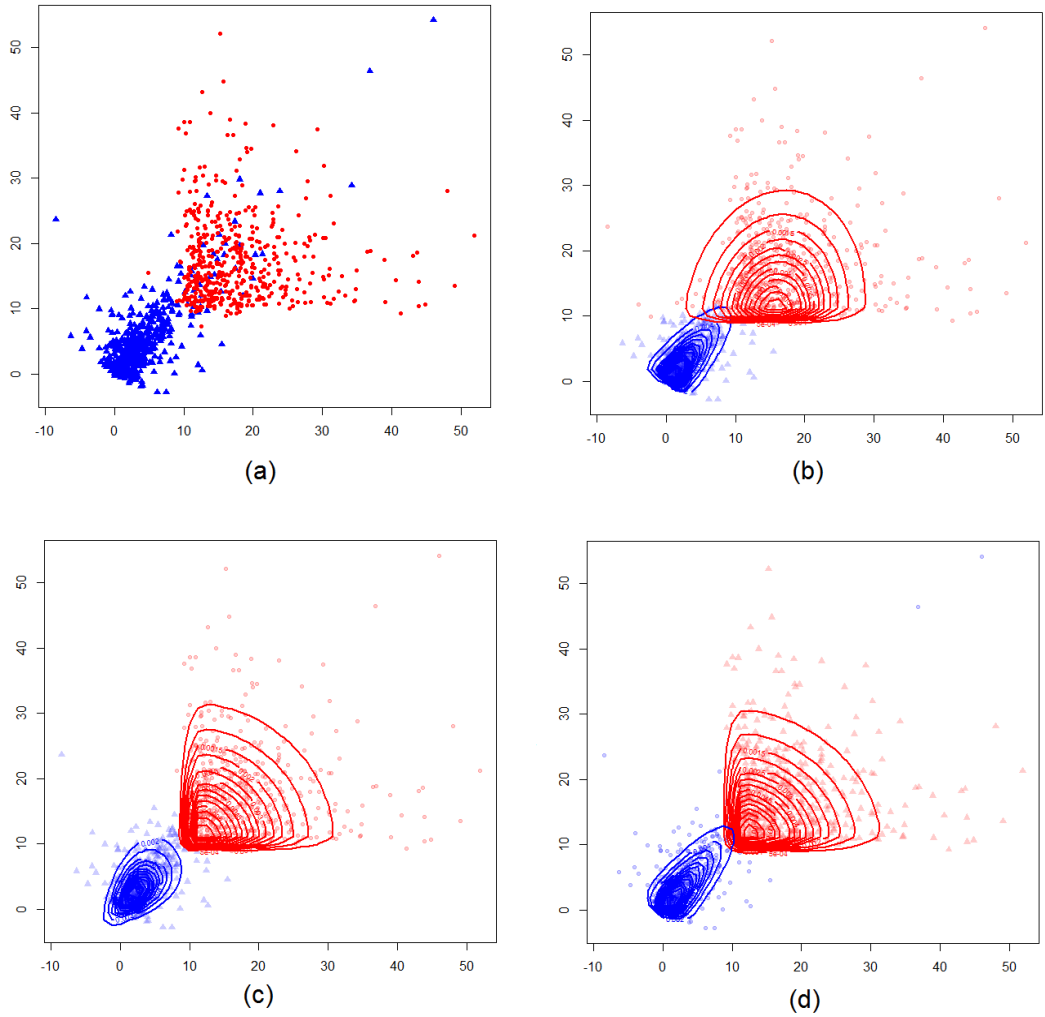

Figure 1: Simulated sample from a FM-CFUST mixture model with one rMST component and one uMST component. (a) Scatter plot of the simulated data with color indicating true cluster labels. (b) The fit given by a FM-rMST distribution, with the density contours of each component overlaid on the scatter plot of the data; (c) The density contours of each component of the fit given by a FM-uMST distribution; (d) Contour plot of the component densities of the fitted FM-CFUST model.

For the upper cluster of data corresponding to those generated from the uMST component (shown as blue triangle in Figure 1), the difference between these models is appreciably wider. It is evident that the FM-uMST and FM-CFUST models provide much closer fits to this cluster of the simulated data than the fit given by the FM-rMST model. Both the FM-uMST and FM-CFUST models produce shell-shaped contours, while that given by the FM-rMST model appears to be shaped like a semi-circle. It is of interest to note that the rMST distribution has difficulty in modelling this type of shape. Again this is not surprising given that the data were generated from a uMST distribution. Also, it is again encouraging that the more general FM-CFUST model is able to model the data accordingly without prior knowledge of the constrains on $\mathbf{\Delta}_h$ under which the data were generated.

To see how the clustering procedures based on the FM-rMST and FM-uMST models perform on data generated from the more general FM-CFUST model, we reported in Table 1 the misclassification rate (MCR) and adjusted Rand index (ARI) (Hubert and Arabie, 1985) for each model. Here, the MCR is calculated for each permutation of the cluster labels of the clustering result under consideration and the rate reported is the minimum value over all such permutations. The ARI is an adjusted version of the Rand index (RI) that corrects for chance agreement, having an expected value of zero if the two clustering results were independent. It takes the maximum value of one when there is perfect agreement between the two clustering results. A higher ARI value corresponds to a closer match between the two clustering results.

As can be expected, it can be observed from Table 1 that the FM-CFUST model has a very good clustering performance, with a lower MCR and higher ARI compared to its two submodels in this example. This is also supported by the visual message of Figure 1(d), where the contours provide a much closer fit to the data. It is also favoured by BIC. Concerning the two nested models, the FM-uMST model is preferred over the FM-rMST model for this dataset according to BIC. The former also achieves a lower MCR and higher ARI compared to the latter model. It should be noted that since the number of unknown parameters is the same for the FM-rMST and FM-uMST models, BIC will favour the model with the larger likelihood value for these two models.

Table 1: Clustering performance of the FM-rMST, FM-uMST, and FM-CFUST distributions on the simulated CFUST mixture data.

| Model | $\log L$ | BIC | MCR | ARI |
|---|---|---|---|---|
| FM-rMST | -6264.44 | 12646.32 | 0.063 | 0.76 |
| FM-uMST | -6248.38 | 12614.18 | 0.049 | 0.81 |
| FM-CFUST | -6195.13 | 12521.50 | 0.044 | 0.83 |

## 4.2 Simulation of independent skew $t$ observations

It is of interest to compare the performance of the three skew $t$ models in handling observations generated from independent (univariate) skew $t$-distributions (the rMST, uMST, and CFUST distributions coincide in the univariate case). Note that neither the rMST, uMST, and CFUST distributions will be equal to the joint distribution of independent variables that have a MST distribution. However, it can be seen from the convolution-type definition of the uMST distribution that it should be able to handle independent skew variables. As noted by Sahu et al. (2003, 2009), the correlations between variables obeying a uMST distribution cannot be all zero. To see this, we have from Sahu et al. (2009) that the scale matrix for the uMST distribution can be expressed as

$$\text{cov}(\boldsymbol{Y}) = \left(\frac{\nu}{\nu-2}\right)\left[\boldsymbol{\Sigma} + \left(1 - \frac{2}{\pi}\right)\boldsymbol{\Delta}^2\right] + a(\nu)\boldsymbol{\delta}\boldsymbol{\delta}^T, \tag{28}$$

where

$$a(\nu) = \frac{2}{\pi}\left(\frac{\nu}{\nu-2}\right) - \frac{\nu}{\pi}\left[\frac{\Gamma\left(\frac{\nu-1}{2}\right)}{\Gamma\left(\frac{\nu}{2}\right)}\right]^2. \tag{29}$$

Table 2: Estimates given by the FM-rMST, FM-uMST, and FM-CFUST distributions on the simulated CFUST mixture data.

|  | true | FM-rMST | FM-uMST | FM-CFUST |
|---|---|---|---|---|
| $\mu_{1,1}$ | 0 | 0.11 | 0.85 | -0.06 |
| $\mu_{1,2}$ | 0 | -0.07 | 0.84 | -0.69 |
| $\mu_{2,1}$ | 10 | 15.66 | 9.92 | 10.49 |
| $\mu_{2,2}$ | 10 | 9.70 | 9.74 | 9.86 |
| $\Sigma_{1,11}$ | 1.5 | 1.37 | 3.02 | 1.66 |
| $\Sigma_{1,12}$ | -1.0 | -0.98 | 3.00 | -1.11 |
| $\Sigma_{1,22}$ | 0.7 | 0.73 | 3.21 | 0.81 |
| $\Sigma_{2,11}$ | 1 | 29.83 | 0.69 | 0.64 |
| $\Sigma_{2,12}$ | 0 | 0.61 | 0.10 | 0.02 |
| $\Sigma_{2,22}$ | 1 | 2.54 | 0.25 | 0.48 |
| $\Delta_{1,11}$ | 4 | 3.30 | 2.48 | 3.84 |
| $\Delta_{1,12}$ | 0 | – | – | 0.40 |
| $\Delta_{1,21}$ | 5 | 4.24 | – | 4.95 |
| $\Delta_{1,22}$ | 0 | – | 3.49 | 1.07 |
| $\Delta_{2,11}$ | 10 | 0.58 | 8.32 | 8.22 |
| $\Delta_{2,12}$ | 0 | – | – | -0.39 |
| $\Delta_{2,21}$ | 0 | 8.18 | – | 0.02 |
| $\Delta_{2,22}$ | 10 | – | 8.70 | 8.15 |
| $\nu_1$ | 3 | 3.02 | 4.61 | 2.89 |
| $\nu_2$ | 3 | 4.64 | 5.86 | 5.49 |
| $\pi_1$ | 0.5 | 0.47 | 0.47 | 0.49 |

It follows that the scale matrix (28) in the skew case ($\boldsymbol{\delta} \neq \boldsymbol{0}$) will be diagonal only if $\boldsymbol{\Sigma}$ is diagonal and, either $\boldsymbol{\delta}\boldsymbol{\delta}^T$ is diagonal (for example, $\boldsymbol{\delta}$ having only one non-zero element) or the term $a(\nu)$ is zero. It can be seen from Figure 2 that this term converges quite quickly to zero as $\nu$ increases, being close to zero for $\nu \geq 10$. It can also be shown formally that $a(\nu)$ tends to zero using the result that

$$\lim_{\nu \to \infty} \frac{\Gamma(\nu+\alpha)}{\Gamma(\nu)\nu^\alpha} = 1, \tag{30}$$

for $\alpha \in \mathbb{R}$ (Wendel, 1948).

Figure 3 shows the scatterplot of 500 observations generated from two independent skew $t$-distributions, where all parameters are taken to be identical (zero location and unit scale) for both variables except for the skewness parameter. Both variables have 3 degrees of freedom. The skewness parameter of the second variable is doubled that of the first variable, given by 5 and 10 respectively. More formally,

$$Y_1 \sim ST(0,1,5,3),$$
$$Y_2 \sim ST(0,1,10,3).$$

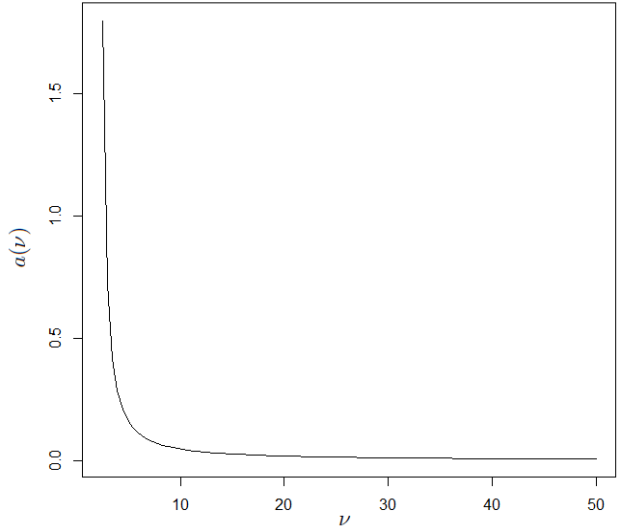

Figure 2: The function $a(\nu)$ versus $\nu$.

Table 3: log likelihood ($\log L$), BIC and KL distance of the rMST, uMST, and CFUST models on the two bivariate simulated independent datasets. The second and third columns correspond to the first set of simulated data. The last two columns correspond to the second set of simulated data.

| Model | $\log L$ | BIC | KL | $\log L$ | BIC | KL |
|---|---|---|---|---|---|---|
| rMST | -3266.59 | 6582.89 | 0.2890 | -3189.61 | 6428.94 | 0.2042 |
| uMST | -3181.36 | 6412.44 | 0.0568 | -3120.07 | 6289.86 | 0.0416 |
| CFUST | -3177.82 | 6411.58 | 0.0405 | -3115.91 | 6287.76 | 0.0367 |

The density contours of the fitted rMST, uMST, and CFUST distributions are depicted in Figure 3(b) to (d), respectively. A visual comparison of the plots shows that both the uMST and CFUST models provide reasonable fits, whereas the rMST model appears to have difficulty in accommodating the skewness that is in the direction of either axes. In particular, this model attempts to capture skewness along the direction of one axis only, as manifested by the steep contours parallel to the vertical axis given by the rMST model in Figure 3 (b). To gain a better insight into the behaviour of these models in this situation, we generated different sets of simulated observations using the same two variables. Interestingly, we observed that the rMST model appears to produce one of two types of results as shown in Figure 3(b) and (f). In some instances, we obtain a fit similar to the first case as depicted in Figure 3(b). For other instances (see Figure 3(f)), the shape of the contours appear to be rotated 90 degrees compared to the first case, where now the steep contours are concentrated along the horizontal axis. On the other hand, the results given by the uMST and CFUST models are quite consistent, as can be observed from Figure 3(c), (d), (g), and (h).

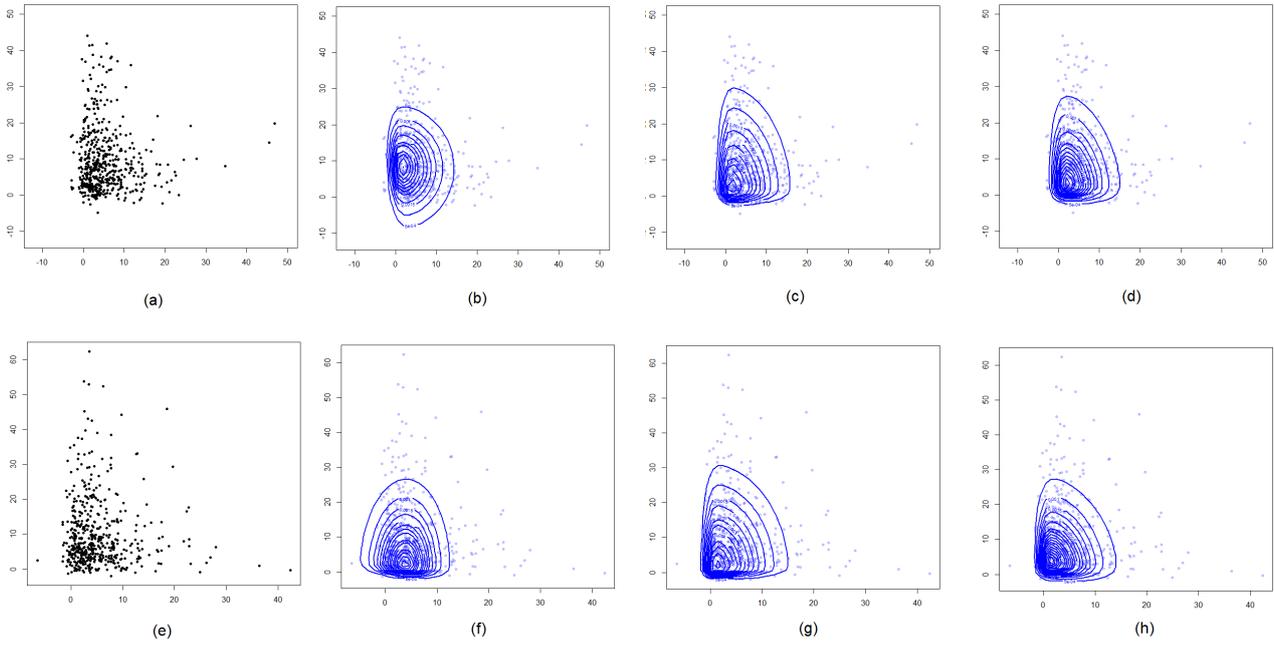

Figure 3: Simulated samples from two independent skew $t$- distribution. First row: first set of simulated data. (a) Scatter plot of the bivariate simulated data; (b) the fit given by a rMST distribution; (c) the density contours given by a fitted uMST distribution; (d) contour plot of the fitted FM-CFUST model. Second row: second set of simulated data. (e) scatter plot of the trivariate simulated data; (f) the fit given by a rMST distribution; (g) the fit given by a uMST distribution; (h) the fit given by a CFUST model.

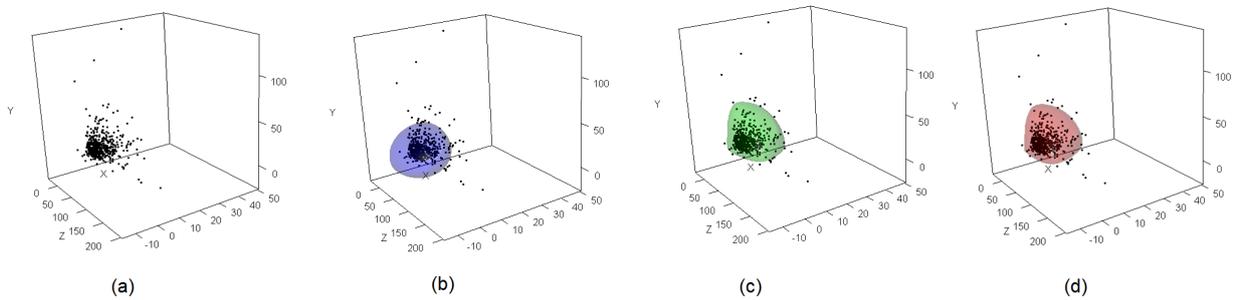

Figure 4: Simulated samples from three independent skew $t$- distribution. (a) Scatter plot of the trivariate simulated data; (b) the fit given by a rMST distribution; (c) the fit given by a uMST distribution; (d) the fit given by a CFUST model.

The above observation is also supported by the estimates of the parameters given by the three fitted models. Focusing first on the estimates of the skewness and scale parameters, they are given by

$$\boldsymbol{\delta}_{rMST} = \begin{bmatrix} 0.23 \\ 10.54 \end{bmatrix}, \boldsymbol{\delta}_{uMST} = \begin{bmatrix} 6.45 \\ 13.22 \end{bmatrix},$$

$$\boldsymbol{\delta}_{CFUST} = \begin{bmatrix} 5.13 & -0.16 \\ -0.81 & 9.84 \end{bmatrix},$$

and

$$\boldsymbol{\Sigma}_{rMST} = \begin{bmatrix} 11.78 & -1.15 \\ -1.15 & 0.94 \end{bmatrix}, \boldsymbol{\Sigma}_{uMST} = \begin{bmatrix} 0.72 & -0.05 \\ -0.05 & 0.68 \end{bmatrix},$$

$$\boldsymbol{\Sigma}_{CFUST} = \begin{bmatrix} 0.97 & 0.24 \\ 0.24 & 2.06 \end{bmatrix},$$

respectively (for the first set of simulated data shown in Figure 3(a)-(d)), it can observed that the first element of $\boldsymbol{\delta}_{rMST}$ is very small compared to the second element, while this is reversed for the diagonal elements of $\boldsymbol{\Sigma}_{rMST}$. This produces semicircle-liked contours with the base parallel to the axis of the second variable. The mode of the fitted rMST model is also not close to the empirical mode of the data, being shifted along the vertical axis in order to accommodate the skewness in the vertical direction. In comparison, the modes of the fitted uMST and CFUST models are closer to the true value. For this dataset, the estimates of the location vector $\boldsymbol{\mu}$ given by the three models are given by

$$\boldsymbol{\mu}_{rMST} = \begin{bmatrix} 3.75 \\ -0.21 \end{bmatrix}, \boldsymbol{\mu}_{uMST} = \begin{bmatrix} -0.47 \\ -0.59 \end{bmatrix},$$

$$\boldsymbol{\mu}_{CFUST} = \begin{bmatrix} 0.00 \\ 1.00 \end{bmatrix}.$$

An analogue of this simulation using three variables is shown in Figure 4. Similar behaviour to the bivariate case can be observed in Figure 4, where the uMST and CFUST models produce similar good fits, while the rMST model does not provide a good fit. To compare the fits quantitatively, we report in Tables 3 and 4 the BIC value and the Kullback-Leibler (KL) distance from the true generating distribution for each of the three models for the bivariate and trivariate simulated data, respectively. It can observed from these tables that the fits given by the uMST and CFUST models are relatively close to each other, with the CFUST model being closer to the true distribution according to the KL distance for the bivariate simulated data, but with the uMST model having a slightly smaller KL distance for the trivariate data. The CFUST model is preferred in terms of BIC for both bivariate and trivariate simulated datasets.

Table 4: $\log L$, BIC and KL distance for the rMST, uMST, and CFUST models on the trivariate simulated independent dataset.

| Model | $\log L$ | BIC | KL |
|---|---|---|---|
| rMST | -5231.60 | 10543.98 | 0.6352 |
| uMST | -5084.86 | 10250.52 | 0.1194 |
| CFUST | -5074.51 | 10248.45 | 0.1196 |

## 4.3 Simulation of CFUST observations

We consider also the fitting of the rMST, uMST, and CFUST models to observations generated from the latter. In particular, these observations were generated from a CFUST distribution that is neither a rMST distribution nor a uMST distribution. Figure 5(a) shows the scatterplot of 1000 bivariate observations generated from a CFUST distribution with contours similar to a shell shape but rotated by approximately 45 degrees anticlockwise. More specifically, they follow a CFUST distribution with parameters given by

$$\boldsymbol{\mu} = \left[\begin{array}{c} 0 \\ 0 \end{array}\right], \boldsymbol{\Sigma} = \left[\begin{array}{cc} 1 & 0 \\ 0 & 1 \end{array}\right], \boldsymbol{\Delta} = \left[\begin{array}{cc} 7 & -11 \\ 7 & 11 \end{array}\right], \nu = 3.$$

Note that this resembles a uMST distribution with the same $\boldsymbol{\mu}$, $\boldsymbol{\Sigma}$ and $\nu$, but with skewness parameter given by $\boldsymbol{\delta} = [10, 15]^T$ and then rotated by approximately 45 degrees anticlockwise.

The contours of the fitted rMST, uMST, and CFUST models are depicted in Figure 5(b) to (d), respectively. It is interesting to observe that both the rMST and uMST model do not provide a good fit, but the fit given by the rMST model is slightly preferred over the uMST model based on the log likelihood, BIC, and KL distance for this simulated data (Table 5).

The estimated parameters for the CFUST distribution are close to the true values, except for the estimated skewness matrix given by

$$\hat{\boldsymbol{\Delta}} = \left[\begin{array}{cc} -9.83 & 6.08 \\ 10.22 & 6.19 \end{array}\right],$$

which may appear to be somewhat different to $\boldsymbol{\Delta}$ at least at first sight. In particular, the columns of $\hat{\boldsymbol{\Delta}}$ would seem to be interchanged with those of $\boldsymbol{\Delta}$. However, it should be noted that the CFUST density is invariant under change of order of the columns in $\boldsymbol{\Delta}$. This can be seen from the stochastic representation in (2). More precisely, the CFUST density is invariant under post-multiplication of $\boldsymbol{\Delta}$ by an orthogonal matrix. It thus has $pq - \frac{1}{2}q(q-1)$ free parameters, which reduces to $\frac{1}{2}p(p+1)$ in the case of $q = p$.

Table 5: BIC values for the fitted models to the simulated CFUST data.

| Model | log$L$ | BIC | KL |
|---|---|---|---|
| rMST | -7559.48 | 15174.23 | 0.2734 |
| uMST | -7608.04 | 15271.33 | 0.3361 |
| CFUST | -7425.61 | 14913.39 | 0.0371 |

# 5 Applications to Real Data

## 5.1 Flow cytometric data

We consider the clustering of a trivariate sample of the Diffuse Large B-cell Lymphoma (DL-BCL) dataset provided by the British Columbia Cancer Agency (Aghaeepour et al., 2013). This sample contains fluorescent intensities of multiple conjugated antibodies (known as markers) stained on a sample of over 3000 cells collected from patients diagnosed with DLBCL. These cells were stained with three markers, namely CD3, CD5, and CD19. As manual experts identified four distinct populations, the task is to gate the cells by clustering the data into four groups. Hence we fit four-component FM-rMST, FM-uMST, and FM-CFUST models to the data.

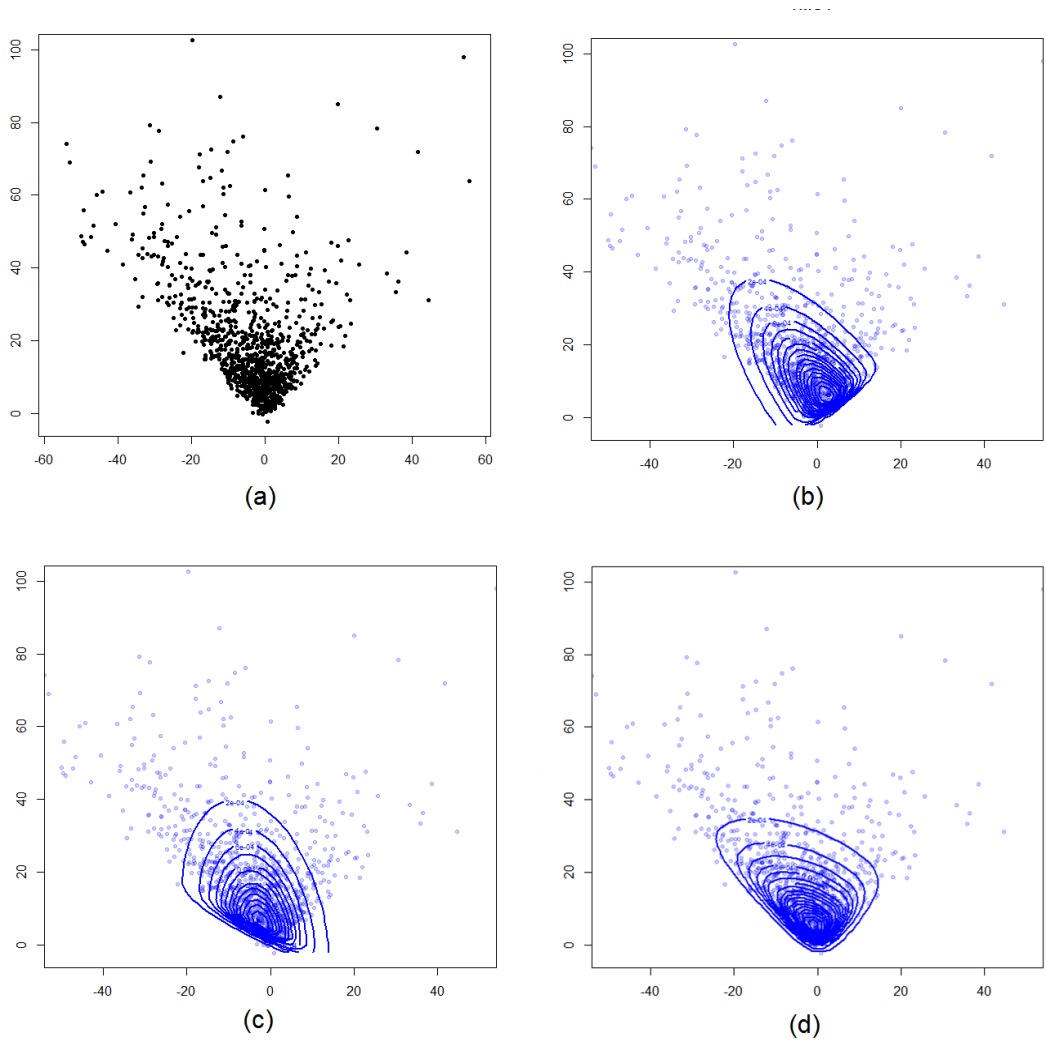

Figure 5: Simulated sample from a CFUST distribution that resembles a uMST distributed rotated approximately 45 degrees anticlockwise. (a) Scatter plot of the simulated data, showing a rotated shell-like shape. (b) The fit given by a rMST distribution; (c) The density contours given by a fitted uMST distribution; (d) Contour plot of the fitted CFUST model.

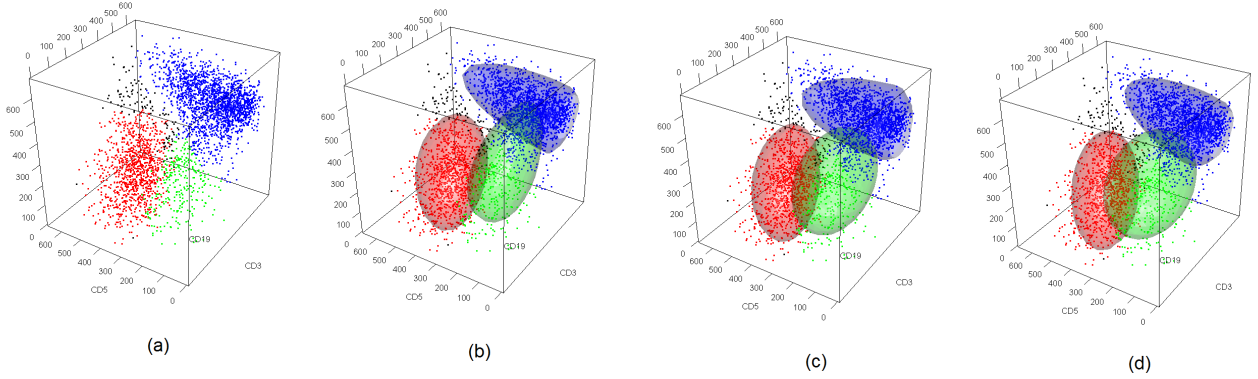

(a) (b) (c) (d)

Figure 6: DLBCL dataset: contour plots of the fitted two- component mixture models on the trivariate data. Scatter plot of LBM and Bfat is given in two colours, red dots for male and blue triangles for female; (a) the fitted mixture contour of the FM-rMST model; (b) the fitted mixture contour of the FM-uMST model; (c) the fitted mixture contour of the FM-CFUST model.

The original data are depicted in Figure 6, where the colours of the dots correspond to the clustering provided by manual gating, which are taken as the 'true' class labels when calculating the MCR for each model. Figure 6 shows the density contours of the components of the fitted FM-rMST, FM-uMST, and FM-CFUST models, respectively, which are displayed with matching colours to Figure 6. To assess the performances of these algorithm on this dataset, we again calculated the MCR and ARI against the 'true' results, with dead cells omitted in the calculation. As can be observed from Figure 6, the fitted contours are very similar. Their log likelihood and BIC values are also quite similar, but with slightly more support for the FM-uMST model. The clustering performance of the latter model is also favoured by the MCR and ARI. From Table 6, it can be observed that the fitted FM-CFUST model under no restrictions on $\boldsymbol{\Delta}$ leads to an inferior result to the FM-uMST model for this dataset. However, as mentioned previously, we would take the solution provided by the FM-uMST model in this situation as this corresponds to a solution with a higher log likelihood and lower BIC value. This illustrates an example where a submodel of the FM-CFUST distribution with diagonal component matrices of skewness parameters $\boldsymbol{\Delta}_h$ can give a preferable solution to the more general version with non-diagonal component matrices $\boldsymbol{\Delta}_h$.

Table 6: Clustering performance of the FM-rMST, FM-uMST, and FM-CFUST distributions on the DLBCL data. *results given by the FM-CFUST corresponding to the highest log likelihood value. †results given by the FM-CFUST without restricting $\boldsymbol{\Delta}$ to be diagonal.

| Model | log$L$ | BIC | MCR | ARI |
|---|---|---|---|---|
| FM-rMST | -59435.10 | 119202.2 | 0.0638 | 0.8100 |
| FM-uMST | -59434.31 | 119200.7 | 0.0405 | 0.8552 |
| FM-CFUST* | -59434.31 | 119200.7 | 0.0405 | 0.8552 |
| FM-CFUST† | -59474.51 | 119354.0 | 0.0575 | 0.8387 |

## 5.2 AIS data

Our final example on real data concerns the Australian Institute of Sport (AIS) data (Cook and Weisberg, 1994), containing 11 biomedical measurements on 202 Australian athletes (100 female and 102 male). For this illustration, we consider the trivariate subset as considered in Lee and McLachlan (2013b), consisting of the variables body mass index (BMI), lean body mass (LBM), and the percentage of body fat (Bfat). We fitted a two-component FM-CFUST model to the data. Again, we consider also the results given by the FM-rMST and FM-uMST models. Figure 7 shows the contours of the fitted density of the FM-rMST, FM-uMST, and FM-CFUST models for each pair of variables of the dataset. Their contours may seem somewhat similar, but subtle differences can be observed, in particular when comparing the cluster corresponding to female (shown as blue triangles in Figure 7) in the second and third rows of Figure 7. The results of the fitted models are tabulated in Table 7. In this example, the FM-CFUST model, which corresponds to the solution with the highest log likelihood, has the smallest MCR and the largest ARI, although the FM-rMST model would be selected using BIC.

Table 7: Clustering performance of the FM-rMST, FM-uMST, and FM-CFUST distributions on the AIS data.

| Model | $\log L$ | BIC | MCR | ARI |
|---|---|---|---|---|
| FM-rMST | -1709.30 | 3561.92 | 0.0347 | 0.866 |
| FM-uMST | -1725.01 | 3593.34 | 0.0198 | 0.922 |
| FM-CFUST | -1700.17 | 3575.51 | 0.0149 | 0.941 |

# 6  Concluding Remarks

We have introduced a finite mixture of CFUST distributions where these distributions include the restricted multivariate skew $t$ and unrestricted multivariate skew $t$-distributions as special cases corresponding to constraints placed on the matrix of skewness parameters. The user has the option of letting the data implicitly choose between the FM-rMST, FM-uMST, and FM-CFUST models via the estimates of the matrix of skewness parameters. Alternatively, the user can make an explicit choice between the three models by using a model selection criterion such as BIC.

Parameter estimation via the EM algorithm has been outlined, following the truncated moments approach in Ho et al. (2012) and Lee and McLachlan (2014a). Furthermore, new results have been derived for the conditional expectation $e_{1hj}^{(k)}$, exploiting an infinite series representation of the logarithm function. This leads to a full implementation of the EM algorithm for the FM-CFUST model with exact and closed-form expressions for all E-step conditional expectations. We have also presented results for five simulations and applications to two real datasets.

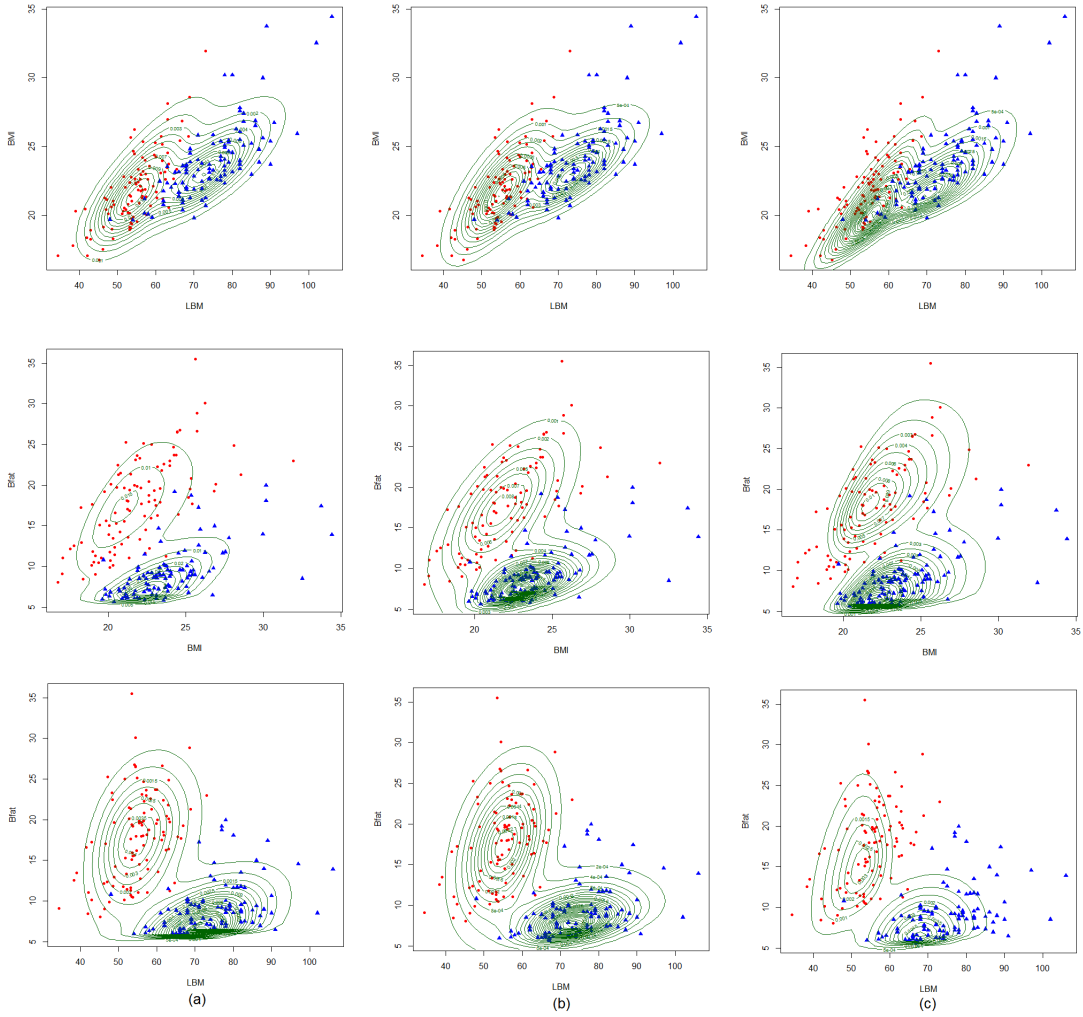

Figure 7: AIS dataset: all pairs of bivariate contour plots of the fitted two-component mixture models on the trivariate data. Scatter plot of LBM and Bfat is given in two colours, red dots for male and blue triangles for female. Different rows correspond to different pairs of variables, while each column corresponds to results provided by a different model. (a) first column: the fitted mixture contour of the FM-rMST model; (b) second column: the fitted mixture contour of the FM-uMST model, (c) third column: the fitted mixture contour of the FM-CFUST model.